\documentclass{aa}

\usepackage[dvips]{graphicx}
\usepackage{txfonts}
\usepackage{natbib}
\usepackage{MR_AA}

\begin{document}

\title{Gravity Darkening in Binary Stars}

\subtitle{}

\author{F. Espinosa Lara\inst{1,2}
\and M. Rieutord\inst{1,2}} 

\institute{Universit\'e de Toulouse; UPS-OMP; IRAP; Toulouse, France
\and CNRS; IRAP; 14, avenue Edouard Belin, F-31400 Toulouse, France}

\date{\today}

\abstract{Interpretation of light curves of many types of binary stars
requires the inclusion of the (cor)relation between surface brightness
and local effective gravity. Until recently, this correlation has
always been modeled by a power law relating the flux or the effective
temperature and the effective gravity, namely $T_\mathrm{eff}\propto
g_\mathrm{eff}^\beta$.}{We look for a simple model that can describe
the variations of the flux at the surface of stars belonging to
a binary system.}{This model assumes that the energy flux is a
divergence-free vector anti-parallel to the effective gravity. The
effective gravity is computed from the Roche model.}{After
explaining in a simple manner the old result of Lucy (1967, Zeit. f\"ur Astrophys. 65,89), which
says that $\beta\sim0.08$ for solar type stars, we first argue that
one-dimensional models should no longer be used to evaluate gravity
darkening laws. We compute the correlation between $\log T_\mathrm{eff}$
and $\log g_\mathrm{eff}$ using a new approach that is valid for
synchronous, weakly magnetized, weakly irradiated binaries.  We show
that this correlation is approximately linear, validating the use of a
power law relation between effective temperature and effective gravity
as a first approximation. We further show that the exponent $\beta$ of
this power law
is a slowly varying function, which we tabulate, of the mass ratio of
the binary star and the Roche lobe filling factor of the stars of the
system. The exponent $\beta$ remains mostly in the interval $[0.20,0.25]$ if
extreme mass ratios are eliminated.}{For binary stars that are
synchronous, weakly magnetized and weakly irradiated, the gravity
darkening exponent is well constrained and may be removed from the free
parameters of the models.}


\maketitle

\section{Introduction}

Gravity darkening refers to the variations of the energy flux at the
surface of a star that result from the variations of the local
effective gravity. This effect is especially visible on rapidly
rotating stars leading to a dark equator and bright poles. It is now a
key effect to interpret the brightness distribution of rapidly rotating
stars when observed by interferometry and to derive
the right
positioning of the rotation axis with respect to the line of sight
\cite[e.g.][]{DKJVONA05,Monnier2007}.

However, gravity darkening also affects the tidally distorted stars in
multiple systems. It has long been known that the interpretation of the
light curves of eclipsing binaries, in particular semi-detached or
contact ones, requires the inclusion of gravity darkening
\cite[e.g.][]{dju03,dju06}. The so-called ellipsoidal variables are
understood as binary stars whose light variations come from the rotation
of their tidally distorted figure \cite[][]{wilson_etal09}. 

The theoretical modelling of gravity darkening started long ago with
the work of \citet{vonzeipel} who discovered that in barotropic
stars the energy flux is proportional to the local effective
gravity leading to the now celebrated von Zeipel law which relates
effective temperature and effective gravity by $T_\mathrm{eff}\propto
g_\mathrm{eff}^{1/4}$. However, this law also assumes a radiative envelope
for the star. Thus, motivated by the importance of gravity darkening
in contact binaries that harbour low mass stars, \citet{lucy67}
studied how to generalize this law to stars with a convective
envelope. Still considering a power-law relation, $T_\mathrm{eff}\propto
g_\mathrm{eff}^\beta$ and calibrating it on one-dimensional models,
Lucy inferred that $\beta\simeq0.08$ for solar-type stars.
 
However, the importance of gravity darkening in stellar
physics motivated further investigations of the theoretical
problem to give better predictions of the exponent $\beta$
\cite[e.g.][]{claret98,claret03,claret12}.  But these works, as
Lucy's one, are based on 1D -models of stellar structure and lead to
gravity darkening laws that depend on the details of the atmosphere
\cite[e.g.][]{claret12}. Moreover, as noted in \cite{ELR11}, these laws
are valid for small deviations from spherical symmetry. It is therefore
not surprising that some observations do not fit because observed gravity
darkening is generally detected on objects that show a strong effect
and are far from spherical.

In \cite{ELR11}, hereafter referred to as paper I, we derived a model
of gravity darkening for rotating stars that is not restricted to
small deviations from sphericity. Moreover, it has no free parameter
and perfectly matches recent observational results on rapidly rotating
stars like $\alpha$~Leo or $\alpha$~Aql. This model rests on very few
assumptions, essentially that the flux is anti-parallel to effective
gravity, and should be very useful for studies of tidally distorted stars.

In this paper we generalize the work of paper~I
to tidally distorted stars in order to provide a robust estimate of the
correlation between flux and local surface gravity. This is achieved
in sections~3 and 4. Before that, we feel that we ought to discuss the
general problem of gravity darkening in spheroidal stars and the previous
attempts to solve it (sect.~2).

\section{The problem of gravity darkening laws in spheroidal stars}

\subsection{Introduction}

Before getting into the details of modelling gravity darkening in tidally
distorted stars, we reconsider the general problem of
deriving a gravity darkening law, namely a law that would relate the
local energy flux and the local effective gravity.

At the surface of a star the energy flux and effective gravity are two
vectors that depend on the colatitude $\theta$ and longitude $\varphi$:

\[ \vF\equiv \vF(\theta,\varphi) \andet
\vg_\mathrm{eff}\equiv\vg_\mathrm{eff}(\theta,\varphi)\]
The existence of a universal relation between these two
quantities is not obvious because they are not locally related by
any physical process (they are not local thermodynamical state
variables!). However, if we consider the star to be barotropic,
that is that the stellar fluid obeys an equation of state where
pressure solely depends on density, namely $P\equiv P(\rho)$,
then all thermodynamical variables only depend on the total
potential (gravitational plus centrifugal) $\Phi$. This implies
that the radiative flux $\vF=-\chi_r\nabla T$ is proportional to
$\vg_\mathrm{eff}=-\nabla\Phi$ because:

\begin{equation} 
\vec F = -\chi_r(\Phi)\nabla T(\Phi) = -\chi_r(\Phi)T'(\Phi)\nabla\Phi=
\chi_r(\Phi)T'(\Phi)\vec g_\mathrm{eff}
\end{equation}
While the surface of a star is generally defined by some value of the opacity, the surface
of a baroropic star can be equivalently defined by an equipotential ($\Phi=\Phi_s$).
The foregoing relation shows that the surface variations of
the flux $\vF$ just follow those of the effective gravity, which is the
famous von Zeipel law.

The assumption of a barotropic fluid is strong because it is enforced
locally. However, the actual thermodynamic structure of a star never
deviates strongly from barotropicity: in convective region matter is
strongly mixed, which homogenizes entropy, while in radiative regions the
recent two-dimensional models of rapidly rotating stars of \cite{ELR11}
show that the relative inclinations of isotherms, isobars or equipotential
is less than half a degree.

In this perspective, the result of \cite{lucy67}, who predicts
$\beta\simeq0.08$, which is a weak dependence of the flux with local
gravity, is quite strange. Moreover, it should apply to stars with a
convective envelope, which is an envelope close to isentropy.

\subsection{Flux-gravity law from spherically symmetric envelopes}

In order to clarify this situation, it is worth trying to understand
the origin of Lucy's result.

Let us first assume that in the range of temperature and density of the
envelope, the opacity of the matter can be approximated by a power law of
the form:

\beq \kappa = \kappa_0 \rho^\mu T^{-s}\eeqn{opapw}
It was noticed by \cite{chandra39} that a spherically symmetric radiative
envelope where both self-gravity and nuclear heat generation can be
neglected behaves like a polytropic gas of index

\beq n = \frac{s+3}{\mu+1}\eeqn{polind}
At the surface of the star, one dimensional models usually assume that
the star radiates like a black body while it is in hydrostatic
equilibrium. These hypothesis are condensed into the boundary condition

\beq P = \frac{2}{3}\frac{g}{\kappabar}\eeqn{bcp}
which is largely used in stellar evolution codes
\cite[e.g.][]{KW98,HK94}. In this expression $g$ is the surface gravity,
$P$ the surface pressure and $\kappabar$ an averaged opacity. We note
that this condition is used for both types of envelopes, convective or
radiative. This may be surprising in the case of a convective envelope
but we recall that in the uppermost layers of a convective zone, convection
is very inefficient and flux is almost only radiative.

Since the atmosphere is polytropic 

\[ P\propto T^{n+1} \andet \rho\propto T^n\; .\]
Thus we find:

\[ g \propto T^{n(\mu+1)+1-s}\]
leading to the gravity darkening exponent:

\beq \beta= \frac{1}{n(\mu+1)+1-s}\eeqn{beta}
We immediately observe that the expression of the polytropic index
\eq{polind} implies that for a radiative envelope

\[ \beta= \frac{1}{4}\]
showing that this exponent is actually independent of the exponents of
the opacity and gives back the von Zeipel law, as expected.

\begin{figure}
\centerline{\includegraphics[height=5cm]{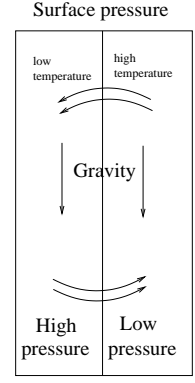}}
\caption[]{Schematic representation of the generation of a baroclinic flow:
the two columns represent two columns of fluid staying at two different
latitudes; the difference in the temperature profile and the assumption
of hydrostatic equilibrium in the column leads to two different pressure
at their base: the fluid flows then from the high pressure region to the
low-pressure one. This circulation transports heat in the latitudinal
direction. In the rotating background of a star, the Coriolis acceleration
transforms this flow into a differential rotation (the so-called thermal
wind in the Earth atmosphere).
} \label{schema}
\end{figure}

We now turn to the convective envelope. In this case the fluid is assumed
almost isentropic and therefore the relation between thermodynamic
variables $P, \rho, T$ is enforced by the $n=3/2$ polytropic index,
which therefore no longer depends on the opacity. However, the
condition \eq{bcp} is still enforced and the opacity may still be
described (locally) by power laws. Thus expression \eq{beta} can be
used with a polytropic exponent $n=3/2$ and the local exponent of the
opacity laws.  At the surface of the Sun, the fitting formula proposed by
J.~Christensen-Dalsgaard for simple solar models \cite[e.g.][]{CDGR95},
suggests $\mu=0.407895$ and $s=-9.28289$ thus leading to

\[ \beta \simeq 0.0807 \]
which is just the value found by \cite{lucy67} for solar type stars.

This result therefore explains the low values that have been derived from
one-dimensional spherical models of solar type stars and shows why the
exponent is sensitive to the details of the opacity in the envelope.

\subsection{The case of non-spherical stars}

Until recently, the lack of multi-dimensionnal models able to describe
stars distorted by rotation or tidal forces, lead to gravity darkening
laws derived from one-dimensional models \cite[][]{claret98,claret12}.
This expedient is built on the assumption that \eq{bcp} is still true
in these spheroidal stars, which is dubious since \eq{bcp} rests on the
hydrostatic equilibrium, namely

\[ \dntau{P}=\frac{g}{\kappa}\; .\]
In rotating or tidally distorted stars, the mechanical balance near the
surface is rather

\[ \vv\cdot\na\vv = -\frac{1}{\rho}\na P -\na\Phi\]
where $\vv$ is the fluid velocity in some inertial frame.
The new velocity terms disappear at the pole of the star but are
important in the equatorial region. Thus, the mechanical balance now
includes the velocity fields that are generated by the baroclinic
torque \cite[e.g.][]{R06,ELR07}. The juxtaposition of 1D-solutions is
not mechanically possible because it generates a latitudinal pressure
desequilibrium (Fig.~\ref{schema} sketches out the mechanism leading to
baroclinic flows). The derivation of a gravity darkening law should
therefore take into account the bi- or tri-dimensionality of the
models.

\section{The model}
\label{sec_model}

\subsection{Introduction}

In order to circumvent the use of one-dimensional models, we propose to
consider the constraints met by the flux in rotating or tidally distorted
stars. First, we know that in the envelope of a star in a steady state

\beq \Div\vF = 0 \eeqn{fluxcons}
i.e. the energy flux $\vF$ is conserved. Second, we may assume that the flux
is almost anti-parallel to the local gravity. We have checked on 2D
models that this assumption is likely very close to reality in isolated
rotating stars, even in the most extreme case. In paper~I, we have shown
that the angle between the two vectors does not exceed half a degree,
which is negligible.

In tidally distorted stars, synchronously corotating with the orbital
motion, this may well be the case because rotation is usually weaker
than in a star rotating near criticality. However, irradiance by the
companion may break this assumption.  Obviously, the additional flux
from the outside may have a dramatic influence if strong enough. Thus,
the following theory applies to binary stars when the companion is weak
enough not to perturb the thermal balance of the primary; other
constraints possibly affecting low-mass stars are discussed in
the conclusion section. To further
simplify the problem we will also assume that the mass distribution of
the primary is like a point mass. This is the usual Roche model,
which is not a strong approximation since we are focussing on the gravity
field in the envelopes.

Provided these hypothesis, we can derive the dependence of the flux
and the effective gravity with latitude and longitude. As we shall see
these quantities are correlated (but not related) as expected from the
previous considerations. To make this correlation easily usable, we will
approximate it by a power law, namely $T_\mathrm{eff}\propto
g_\mathrm{eff}^\beta$, but this has no physical significance except of
showing the importance of the discrepancy with von Zeipel law.

\subsection{Equations}

The main hypothesis is the same described in paper I. We
consider that the energy is transported predominantly in the vertical
direction indicated by the local effective gravity $\vec g_\mathrm{eff}$,
that is, the flux vector $\vec F$ is always anti-parallel to the
effective gravity

\begin{equation}
\label{eq_hyp}
\vec F=-f(r,\theta,\varphi)\vec g_\mathrm{eff}\;,
\end{equation}
where $f$ is an unknown positive function of the position to be determined
and $r$, $\theta$ and $\varphi$ are the spherical coordinates. Combining
this relation with flux conservation \eq{fluxcons}, we get

\begin{equation}
\label{eq_diff}
\vec g_\mathrm{eff}\cdot\nabla f+f\nabla\cdot\vec g_\mathrm{eff}=0\;.
\end{equation}
As in paper I, we will consider for simplicity that the star
is centrally condensed and adopt the Roche model for the effective
gravity. Then, all the energy is generated at the centre of the star,
which implies that

\begin{equation}
\label{eq_bc}
\lim_{r\rightarrow 0} f=f_0=\frac{L}{4\pi GM}\;.
\end{equation}
If we denote by $\vec n$ the unit vector in the vertical
direction, such that $\vec g_\mathrm{eff}=-g_\mathrm{eff}\vec n$, then
we can write

\begin{equation}
\label{eq_grad}
\vn\cdot\nabla\log f=\frac{\nabla\cdot\vg_\mathrm{eff}}{g_\mathrm{eff}}\;,
\end{equation}
Eq. (\ref{eq_grad}) gives the variation of $f$ in the local vertical
direction.  Let us build a trajectory $\mathcal{C}(\theta_0,\varphi_0)$
that starts at the centre of the star with an initial direction
indicated by $(\theta_0,\varphi_0)$ and is tangent to $\vec n$ at every
point. $\mathcal{C}(\theta_0,\varphi_0)$ is therefore a field line of
the effective gravity field. The value of $f$ at a point $\vr$ along the
curve can be calculated as a line integral

\begin{equation}
\label{eq_int}
f(\vec r)=f_0\exp\left(\int_{\mathcal{C}(\theta_0,\varphi_0)}
\!\!\!\!\!\frac{\nabla\cdot\vec g_\mathrm{eff}}{g_\mathrm{eff}}\,\mathrm{d}l\right)
\qquad\mbox{for } \vec r \in \mathcal{C}(\theta_0,\varphi_0)\;.
\end{equation}

According to the Roche model, the effective potential of a
synchronized binary system with circular orbits is given by

\begin{equation}
\begin{array}{rl}
\phi=&\displaystyle -\frac{GM_1}{r}-\frac{GM_2}{\sqrt{a^2+r^2-2ar\cos\theta}}\\
&\displaystyle -\frac{1}{2}\Omega^2r^2(\sin^2\theta\sin^2\varphi+\cos^2\theta)
+a\frac{M_2}{M_1+M_2}\Omega^2r\cos\theta\;,
\end{array}
\end{equation}
where $a$ is the distance between the centres of the two stars and $M_1$
and $M_2$ are the respective masses. The coordinate system whose origin
is at the centre of the primary star, is 
oriented so that the $z$-axis corresponds to the
axis connecting the stars centres. The plane $Oyz$ is the orbital
plane and the $x$-axis is normal to this plane. This implies that the
spherical angles are not the usual colatitude and longitude of the
star. The orbital angular velocity is

\begin{equation}
\Omega=\sqrt{\frac{G(M_1+M_2)}{a^3}}\;.
\end{equation}
We normalise the expression of the potential $\phi$ by $GM_1/a$, so
that

\begin{equation}
\begin{array}{rl}
\tilde\phi=&\displaystyle -\frac{1}{\tilde r}-\frac{q}{\sqrt{1+\tilde r^2-2\tilde r\cos\theta}} \\
&\displaystyle -\frac{1}{2}(1+q)\tilde r^2(\sin^2\theta\sin^2\varphi+\cos^2\theta)+q\tilde r\cos\theta\;,
\end{array}
\end{equation}
where $\tilde \phi=\frac{a}{GM_1}\phi$ is the normalised potential and
$\tilde r=r/a$, the normalised radius. We have also introduced the mass ratio

\begin{equation}
q=\frac{M_2}{M_1}.
\end{equation} 
Note that the normalised angular velocity is $\omega=\sqrt{1+q}$. The
effective gravity is $\tilde{\vec g}_\mathrm{eff}=-\nabla\tilde\phi$
and satisfies

\begin{equation}
\label{eq_div_roche}
\nabla\cdot\tilde{\vec g}_\mathrm{eff}=2(1+q)
\end{equation}

\subsection{Solution of the model}

\begin{figure}
\resizebox{\hsize}{!}{\includegraphics{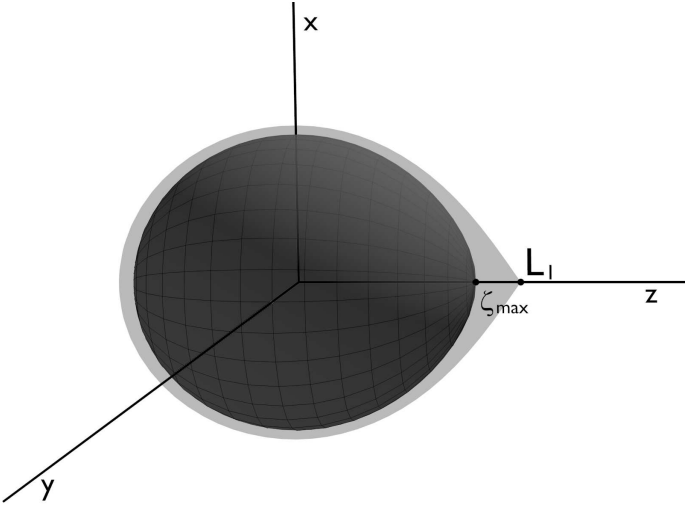}}
\caption{Schematic representation of the primary star with filling
factor 0.8.  The position of the Lagrange point $L_1$is indicated.}
\label{fig_binary}
\end{figure}

We limit ourselves to the case of detached or semi-detached binaries,
focusing on one of the stars that we call the primary. Unfortunately,
eq. (\ref{eq_int}) has resisted all our efforts of finding a solution
in a closed form forcing us to solve it by numerical integration
in the following way.

Let $\zeta$ be the $z$-coordinate of
the point at which an equipotential surface $\tilde{\phi}=$const. intersects
the axis connecting the two stars, normalised by $a$ the position of
the secondary.  Its maximum value $\zeta_\mathrm{max}$ corresponds to
the surface of the primary. If $\zeta_1$ is the distance from the centre of the
primary to the Lagrange point $L_1$ we define the filling factor as
$\rho=\zeta_\mathrm{max}/\zeta_1$ (see Fig. \ref{fig_binary}).
 A star with $\rho=1$ fills its Roche lobe.
This definition is useful because, apart from normalising constants, the
solution depends only on the mass ratio $q$ and the filling factor $\rho$.

As $\zeta$ marks a single equipotential, the potential can be
expressed as a function of $\zeta$ only, $\phi=\phi(\zeta)$ and

\begin{equation}
\label{eq_curv1}
\mathrm{d}\phi=g_0(\zeta)\mathrm{d}\zeta=-\vec g_\mathrm{eff}\cdot\mathrm{d}\vec r\;,
\end{equation}
where $g_0(\zeta)=||\vec g_\mathrm{eff}(x=0,y=0,z=\zeta)||$.

We can use $\zeta$ as the parameter describing the field line
$\mathcal{C}(\theta_0,\varphi_0)$ that appears in (\ref{eq_int}). The
differential equation of the field line is given by

\begin{equation}
\label{eq_curv2}
\vec g_\mathrm{eff}\times\mathrm{d}\vec r=\vec 0\;.
\end{equation}
which expresses that the differential vector is parallel to the effective
gravity. Then, using (\ref{eq_curv1}) and (\ref{eq_curv2}), we can write

\begin{equation}
\label{eq_curve}
\frac{\mathrm{d}\vec r}{\mathrm d\zeta}=\frac{-g_0}{g_\mathrm{eff}^2}\vec g_\mathrm{eff}\;,
\end{equation}
from which we can calculate the field line points
$\vr(\zeta,\theta_0,\varphi_0)$,
for a given starting
direction $(\theta_0,\varphi_0)$. The line element is

\begin{equation}
\label{eq_dl}
\mathrm{d} l=||\mathrm{d}\vec r||=\frac{g_0}{g_\mathrm{eff}}\mathrm{d}\zeta\;.
\end{equation}

Combining ($\ref{eq_int}$), ($\ref{eq_div_roche}$) and
($\ref{eq_dl}$), we get the function $f$ which relates the energy flux
and the effective gravity. We find

\begin{equation}
\label{eq_f}
f(\theta_s,\varphi_s)=f_0\exp\left(\int_0^{\zeta_\mathrm{max}}
\frac{2(1+q)}{\left[g_\mathrm{eff}(\vec r(\zeta,\theta_0,\varphi_0))\right]^2}g_0(\zeta)\mathrm{d}\zeta\right)\;.
\end{equation} 
where $\theta_s$ and $\varphi_s$ are the coordinates of the point where
the field line $\mathcal{C}$ intersects the surface of the primary.

Equations \eq{eq_grad} and (\ref{eq_curve}) form a system of four
differential equations for solving $f$ that is well-suited for
numerical integration. Given an initial direction $(\theta_0,\varphi_0)$
and using initial condition (\ref{eq_bc}), the system can be
integrated from the centre to the surface. We thus get the value of $f$
at the surface for each $(\theta_0,\varphi_0)$ and thus for each
$(\theta_s,\varphi_s)$. There is a one-to-one correspondance between
these coordinates because these field lines do not cross and effective
gravity is defined everywhere.

The integration has been performed using the Runge-Kutta-Fehlberg method,
often referred to as RKF45, which combines two RK methods of fourth and
fifth order \cite[][]{numrec}. It provides an estimate of the truncation error that can
be used to adapt the step size.

\section{Results}

\begin{figure*}
\centering
\includegraphics[width=17cm]{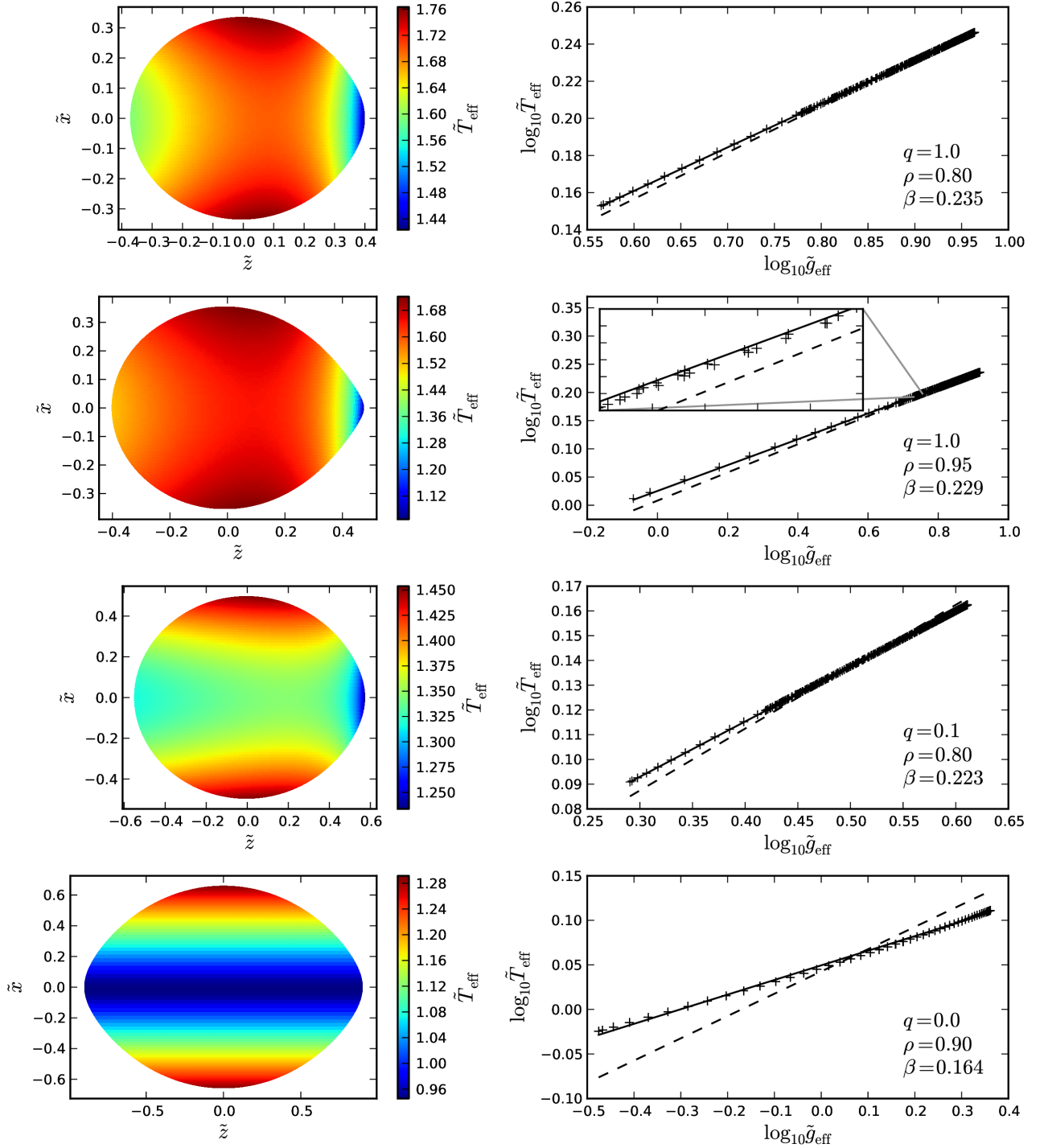}
\caption{\emph{Left}: Distribution of $T_\mathrm{eff}$ at the surface in
one hemisphere of the primary, projected in the plane XZ for several
combinations of the mass ratio $q=M_2/M_1$ and the filling factor
$\rho$. \emph{Right}: Logarithmic plot of $T_\mathrm{eff}$ versus
$g_\mathrm{eff}$ for the same cases. Solid lines are linear fits
to the data and the corresponding value of the gravity darkening
exponent $\beta$ is indicated in the plots. For comparison,
von Zeipel's law ($\beta=0.25$) is shown by a dashed line. In the
second row an inset shows the small dispersion of the points around the
fitting power law.
The last case ($q=0$) corresponds to an isolated star
rotating with angular velocity $\Omega=0.854\Omega_k$.  Normalised
values of effective temperature and gravity are defined in Eq.~\ref{scaled}.}
\label{fig1}
\end{figure*}

Figure \ref{fig1} shows the results for several combinations of the mass
ratio $q$ and filling factor $\rho$. The normalised values of effective
temperature and gravity that appear in the figure are defined as

\begin{equation}
\tilde T_\mathrm{eff}=\left(\frac{4\pi\sigma a^2}{L}\right)^{1/4} T_\mathrm{eff}
\qquad\mbox{and}\qquad
\tilde g_\mathrm{eff}=\frac{a^2}{GM_1} g_\mathrm{eff}\;.
\label{scaled}
\end{equation}
In the left, we can see the distribution of $T_\mathrm{eff}$ at the
surface in one hemisphere of the star, projected in the plane XZ (remember
the definition of the coordinate system shown in Fig. \ref{fig_binary}). A
logarithmic plot of $T_\mathrm{eff}$ versus $g_\mathrm{eff}$ for each
case can be seen on the right. Since the problem has no azimuthal
symmetry, the same value of $g_\mathrm{eff}$ appears at various places
on the surface, but associated with different values of
$T_\mathrm{eff}$, thus illustrating the absence of a one-to-one
relation between $T_\mathrm{eff}$ and $g_\mathrm{eff}$. However, as
shown by the plots, the dispersion is very small (see the inset in
Fig.~\ref{fig1}). It emphasises a good
correlation between the two foregoing quantities. Moreover, the log-log
plot shows that this correlation is well approximated by a power law.
The actual value of the exponent $\beta$ is always slightly less than
1/4. The worst case is the last one where
$q=0$. In fact, this case corresponds to the limit of an isolated star
($M_2=0$) rotating at angular velocity

\begin{equation}
\Omega=\rho^{3/2}\Omega_k\;,
\end{equation}
where $\Omega_k$ is the Keplerian angular velocity at the equator
\mbox{$\Omega_k=\sqrt{\frac{GM_1}{R_e^3}}$}, for which we recover the
results obtained in paper~I.

\begin{table*}
\caption{Fitted values of the gravity darkening exponent $\beta$ in the
correlation between $T_\mathrm{eff}$ and $g_\mathrm{eff}$ as a function
of the mass ratio $q=M_2/M_1$ and the filling factor $\rho$. As the
relation does not follow exactly a power law, the maximum relative error
in $T_\mathrm{eff}$ calculated using the fit compared with the exact
value is included in parentheses.}
\label{table1}
\centering
\begin{tabular}{lp{1.8cm}p{1.8cm}p{1.8cm}p{1.8cm}p{1.8cm}p{1.8cm}p{1.8cm}}
\hline\hline
Mass ratio ($q$)&
\multicolumn{7}{c}{Filling factor ($\rho$)}\\
& $\rho$=0.3  & $\rho$=0.5  & $\rho$=0.7  & $\rho$=0.8  & $\rho$=0.9  & $\rho$=0.95  & $\rho$=0.99  \\[0.1cm]
\hline
$q$=0.0 &
	0.2456\newline (0.00004\%) &
	0.2308\newline (0.0038\%) &
	0.2034\newline (0.078\%) &
	0.1854\newline (0.27\%) &
	0.1641\newline (0.96\%) &
	0.1509\newline (2.08\%) &
	0.1365\newline (6.08\%) \\[0.1cm]
$q$=0.001 &
	0.2464\newline (0.00002\%) &
	0.2342\newline (0.0021\%) &
	0.2110\newline (0.043\%) &
	0.1958\newline (0.15\%) &
	0.1789\newline (0.48\%) &
	0.1708\newline (0.92\%) &
	0.1666\newline (1.55\%) \\[0.1cm]
$q$=0.01 &
	0.2472\newline (0.00002\%) &
	0.2374\newline (0.0013\%) &
	0.2188\newline (0.025\%) &
	0.2068\newline (0.090\%) &
	0.1948\newline (0.31\%) &
	0.1901\newline (0.56\%) &
	0.1880\newline (1.31\%) \\[0.1cm]
$q$=0.1 &
	0.2482\newline (0.00004\%) &
	0.2420\newline (0.0012\%) &
	0.2303\newline (0.014\%) &
	0.2232\newline (0.049\%) &
	0.2168\newline (0.18\%) &
	0.2143\newline (0.36\%) &
	0.2131\newline (0.66\%) \\[0.1cm]
$q$=1.0 &
	0.2489\newline (0.00004\%) &
	0.2452\newline (0.0013\%) &
	0.2385\newline (0.011\%) &
	0.2346\newline (0.050\%) &
	0.2308\newline (0.23\%) &
	0.2291\newline (0.52\%) &
	0.2280\newline (1.44\%) \\[0.1cm]
$q$=3.0 &
	0.2490\newline (0.00004\%) &
	0.2455\newline (0.0015\%) &
	0.2394\newline (0.020\%) &
	0.2359\newline (0.092\%) &
	0.2324\newline (0.37\%) &
	0.2308\newline (0.80\%) &
	0.2297\newline (2.18\%) \\[0.1cm]
$q$=10.0 &
	0.2489\newline (0.00005\%) &
	0.2453\newline (0.0019\%) &
	0.2391\newline (0.036\%) &
	0.2355\newline (0.15\%) &
	0.2320\newline (0.55\%) &
	0.2304\newline (1.16\%) &
	0.2292\newline (3.09\%) \\[0.1cm]
$q$=100.0 &
	0.2487\newline (0.00005\%) &
	0.2447\newline (0.0029\%) &
	0.2377\newline (0.061\%) &
	0.2337\newline (0.23\%) &
	0.2298\newline (0.82\%) &
	0.2279\newline (1.75\%) &
	0.2267\newline (4.57\%) \\[0.1cm]
\hline
\end{tabular}
\end{table*}

In Table \ref{table1} we summarize the fitted values of the gravity
darkening exponent as a function of $q$ and $\rho$. As pointed out
above, the relation between $T_\mathrm{eff}$ an $g_\mathrm{eff}$ is not
exactly a power law so, in order to measure the quality of the fit,
we also included, in parentheses, the maximum relative error
in $T_\mathrm{eff}$ calculated using the fit compared to the exact
value. Again, for intermediate $q$, the departures from von Zeipel's
value are moderate, with a value of $\beta$ above $\sim 0.21$
and a correlation between $\log T_\mathrm{eff}$ and $\log g_\mathrm{eff}$
that remains more or less linear.

\begin{figure}
\resizebox{\hsize}{!}{\includegraphics{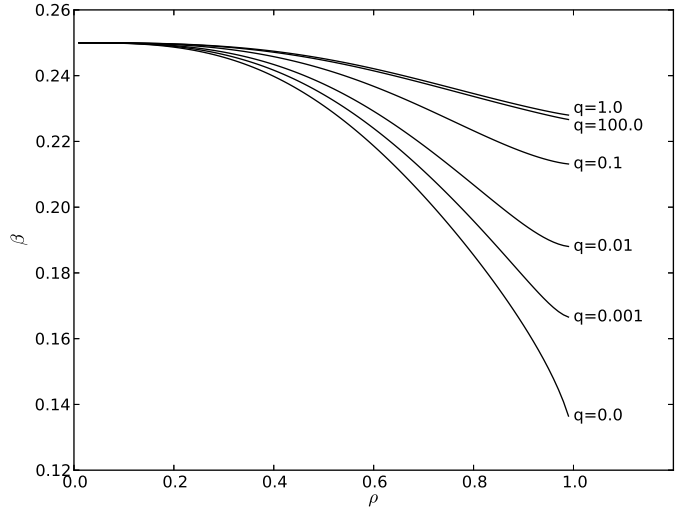}}
\caption{Fitted values of the gravity darkening exponent as a function
of the filling factor $\rho$ for different values of the mass ratio
$q=M_2/M_1$. Von Zeipel's value $\beta=0.25$ is only attained in the
limit $\rho=0$ (spherical stars).}
\label{fig2}
\end{figure}

The largest deviation is observed for small $q$, for which the effects
of nonlinearity are also the strongest. The same behavior can be observed
in Fig.~\ref{fig2}, where we plot the fitted values of $\beta$ versus
the filling factor $\rho$ for different values of $q$. As expected,
the asymptotic limit of $\rho=0$, i.e. spherical stars, corresponds to
von Zeipel's value $\beta=1/4$. For a given $\rho$, the maximum value
of $\beta$ is attained for $q\sim 3$.

\section{Conclusions}

We have explained in a simple way the old result of \cite{lucy67}, who
derived a gravity darkening law with an exponent $\beta\sim0.08$ from
1D models of solar-type stars. We trace this value of $\beta$ as a consequence
of the boundary conditions used by these stellar models, namely that
surface pressure varies like $g/\kappabar$ and that convective envelopes
are close to a $n=3/2$ polytrope.

Noting that 2D or 3D models of rotating or tidally distorted stars cannot
be in hydrostatic equilibrium, we argue that the use of gravity darkening
exponents derived from 1D models should be avoided. Following paper~I,
we propose a simple model of the gravity darkening effect that rests on
hypothesis generally valid in the envelope of stars:

\begin{enumerate}
\item flux is conserved (no heat sources);
\item flux is antiparallel to local effective gravity;
\item gravity is close to that of the Roche model.
\end{enumerate}

These three hypothesis lead to a correlation between effective temperature
and effective surface gravity which can be described by a power law
with a gravity darkening exponent $\beta$ that weakly depends on mass
ratio and Roche lobe filling factor. If we discard extreme mass ratios,
$\beta$ remains in the interval $[0.2, 0.25]$. It is also independent
of the nature (convective or radiative) of the envelope.

This result is interesting to estimate gravity darkening in bright
(hot) components of binary stars. It shows that the assumption of the
von Zeipel value $\beta=1/4$ is certainly too restrictive.

However, a blind use of this model in binary stars is not recommended
either. Indeed, this model does not include:

\renewcommand{\theenumi}{\alph{enumi}}
\begin{enumerate}
\item irradiation from the secondary,
\item Coriolis effects on convection,
\item magnetic fields,
\item possible non-synchronism.
\end{enumerate}

The first two effects may indeed invalidate the assumption of the
antiparallelism between flux and gravity. Magnetic field may generate
spotted regions thus destroying the correlation between effective gravity
and flux. Finally, non-synchronism, which may come from differential
rotation, alters the assumed shape of the star.

The first three effects may especially affect low-mass companions of
binary stars. For these stars, it is likely that the correlation between
surface effective gravity and effective temperature is perturbed. The
results of \cite{dju03,dju06} on semi-detached binaries, which give
small gravity darkening exponents (namely $\beta\sim0.1$) for late-type
companions, may be viewed as the signature of a weak correlation between
$T_\mathrm{eff}$ and $g_\mathrm{eff}$ for these stars.

In conclusion, the gravity darkening exponent should be interpreted
as the parametrization of a correlation between $T_\mathrm{eff}$
and $g_\mathrm{eff}$. For binary star components, where conditions
(a-b-c-d) do not apply, this exponent should be fixed to a value derived
from Table~1 according to the mass-ratio of the system and the filling
factors of the stars. If one ore more effect of the list exists, clearly
a more elaborated model is needed.

\begin{acknowledgements}
The authors acknowledge the support of the French Agence Nationale de
la Recherche (ANR), under grant ESTER (ANR-09-BLAN-0140).  This work
was also supported by the Centre National de la Recherche Scientifique
(CNRS, UMR 5277), through the Programme National de Physique Stellaire
(PNPS). The numerical calculations have been carried out on the CalMip
machine of the `Centre Interuniversitaire de Calcul de Toulouse' (CICT)
which is gratefully acknowledged.
\end{acknowledgements}

\bibliographystyle{aa} 
\bibliography{../../biblio/bibnew} 

\end{document}